\documentclass[journal]{IEEEtran}
\usepackage{graphicx}
\usepackage{caption}
\usepackage{subcaption}
\usepackage{float}

\begin{document}

\title{Precision Muon Tracking Detectors for High-Energy Hadron Colliders}

\author{Ph.~Gadow, O.~Kortner, \underline{H.~Kroha}$^*$\thanks{$^*$Corresponding author: kroha@mpp.mpg.de}, R.~Richter
 \\ \textit{Max-Planck-Institut f\"ur Physik, F\"ohringer Ring 6, D-80805 Munich, Germany}}

\maketitle
\pagestyle{empty}
\thispagestyle{empty}

\begin{abstract}

Small-diameter muon drift tube (sMDT) chambers with 15~mm tube diameter are a cost-effective technology for high-precision muon tracking over large areas at high background rates 
as expected at future high-energy hadron colliders including HL-LHC.
The chamber design and construction procedures have been optimized for mass production and provide sense wire positioning accuracy of better than 10~$\mu$m.    
The rate capability of the sMDT chambers has been extensively tested at the CERN Gamma Irradiation Facility. It exceeds the one of the ATLAS muon drift tube (MDT) chambers, 
which are operated at unprecedentedly high background rates of neutrons and $\gamma$-rays, by an order of magnitude, which is sufficient for almost the whole of the muon detector 
acceptance at FCC-hh at maximum luminosity. sMDT operational and construction experience exists from ATLAS muon spectrometer upgrades which are in progress or under preparation
for LHC Phase 1 and 2.
\end{abstract}


\section{Introduction}

\begin{figure}[h!]
	\centering
	\begin{subfigure}[b]{0.45\textwidth}
        \hspace{-3mm}
        \includegraphics[width=1.1\textwidth]{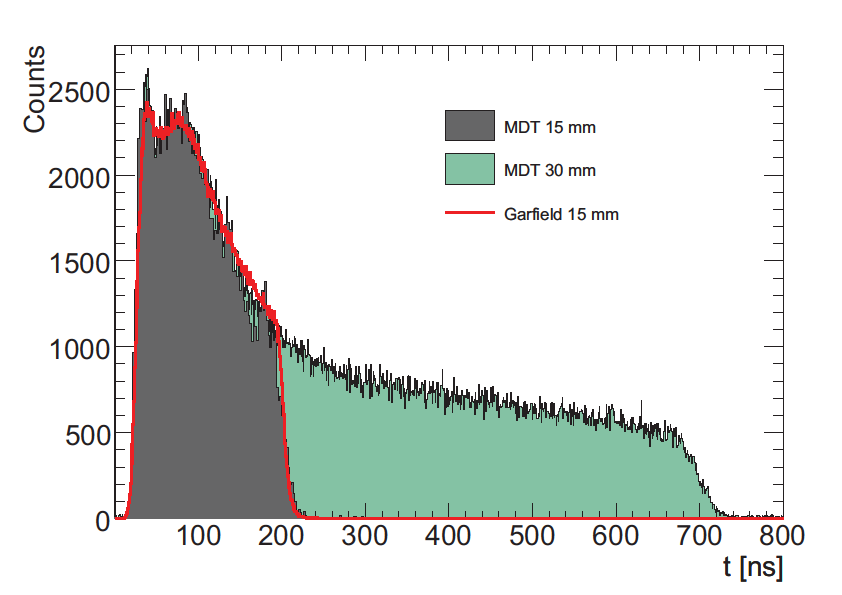}
        \caption{}
	\label{fig::spectrum}
	\end{subfigure}
	\qquad
	\begin{subfigure}[b]{0.45\textwidth}
	\hspace{-2mm}     
         \includegraphics[width=1.1\textwidth]{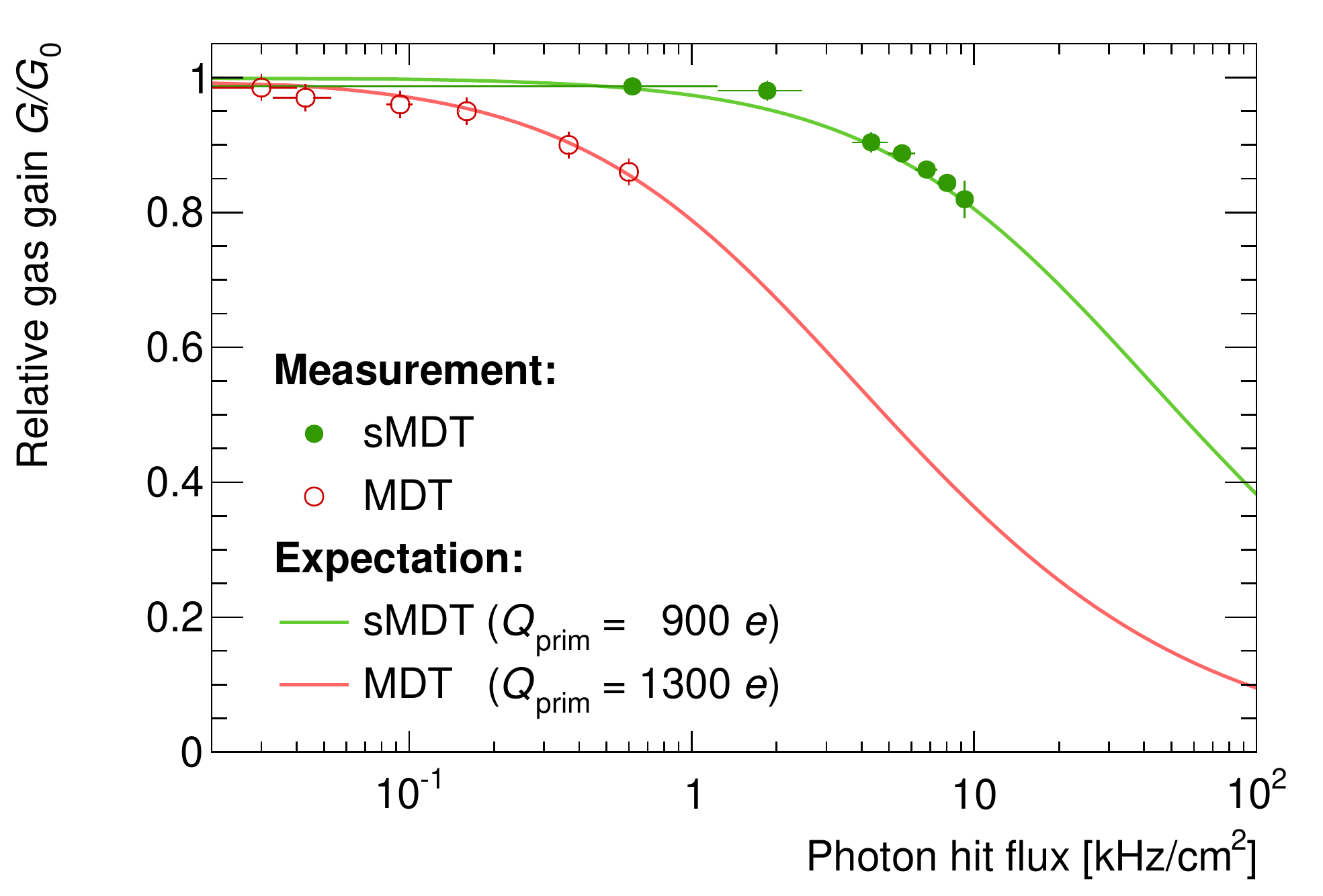}
         \caption{}
         \label{fig::gain}
         \end{subfigure}
\vspace{-3mm}
\caption{(a) Drift time spectra of MDT (light grey) and sMDT tubes (dark grey).
         (b) Gas gain measured as a function of $\gamma$ background flux for MDT (open circles) and sMDT tubes (full circles) compared to expectations.}
\end{figure}

\begin{figure}[h!]
\centering
\includegraphics[width=0.5\textwidth]{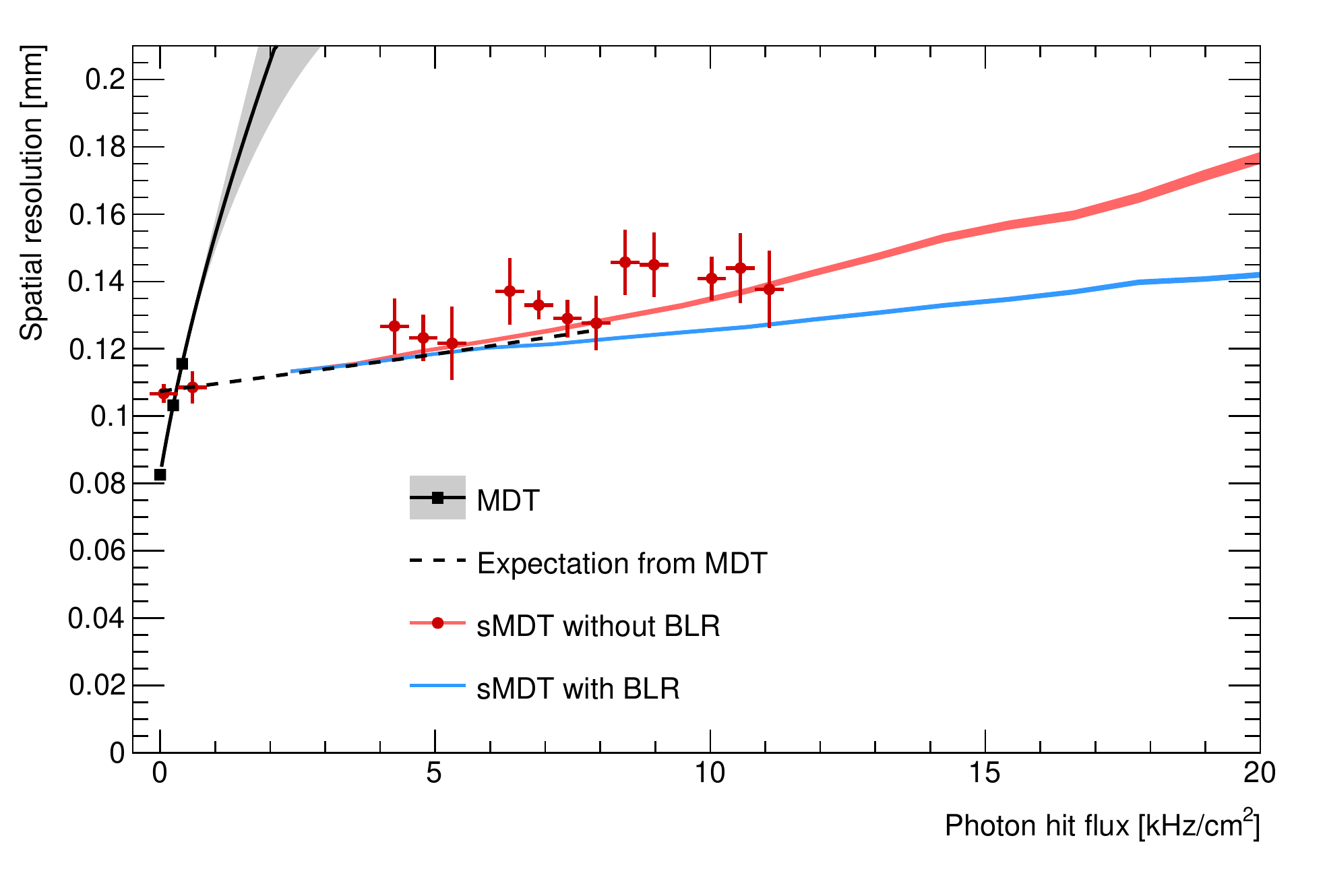}
\caption{Measured average spatial resolution of MDT and sMDT drift tubes as a function of the background flux compared to simulations with and without active 
baseline restoration (BLR) of the readout electronics.} 
\label{fig:resolution}
\end{figure}

\begin{figure}[h!]
\centering
\includegraphics[width=0.5\textwidth]{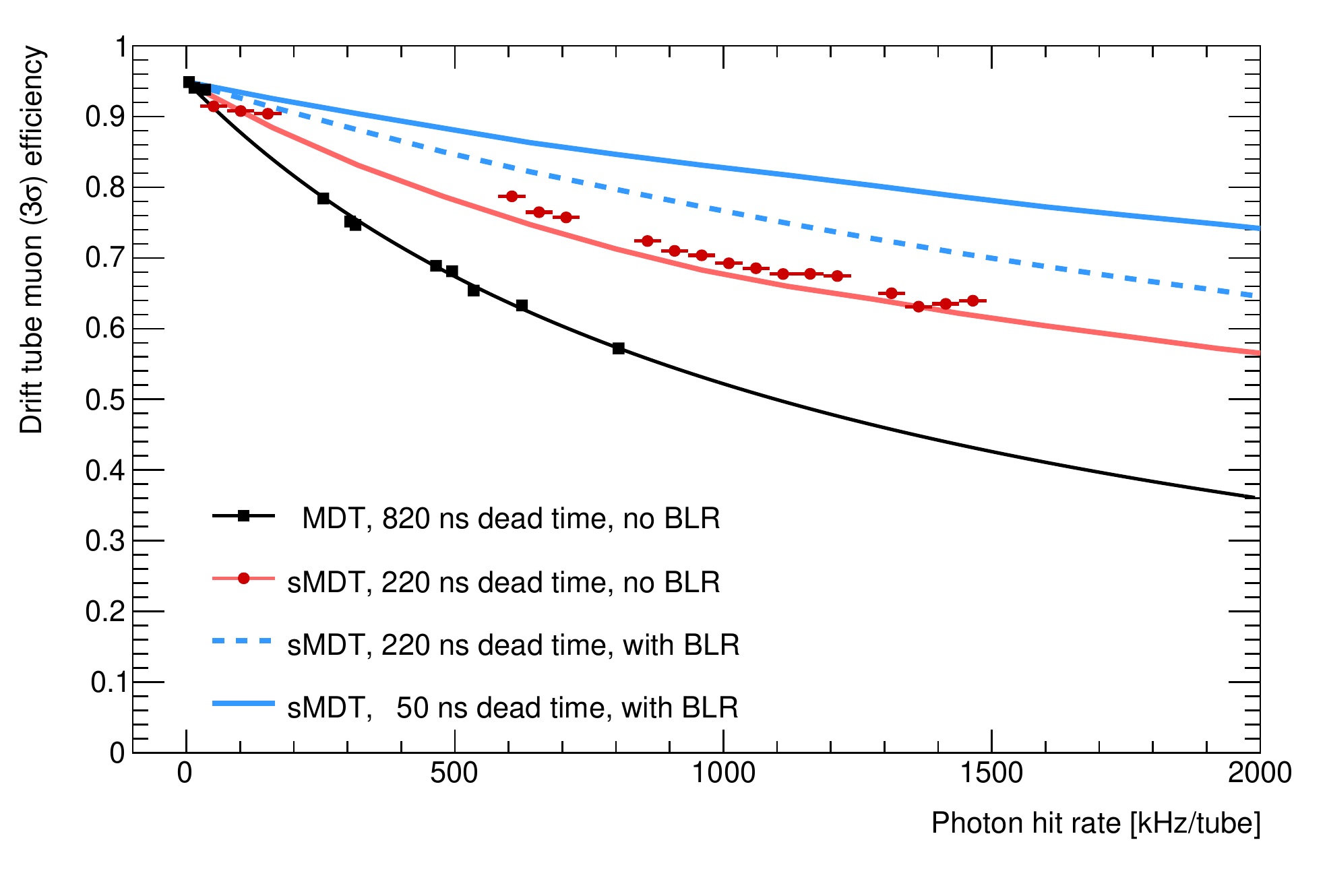}
\caption{Muon detection efficiency of MDT and sMDT drift tubes with different electronics deadtime (see text) as a function of the background hit rate compared 
to simulations with and without active baseline restoration (BLR) of the readout electronics.} 
\label{fig:efficiency}
\end{figure}

Drift tube detectors are a cost-effective technology for high-precision muon tracking over large areas. Monitored drift tube (MDT) chambers with 30~mm diameter drift tubes 
are used for precision muon tracking in the ATLAS muon spectrometer~\cite{ATLAS}. Operated with Ar:CO$_2$ (93:7) drift gas at 3 bar, they provide excellent spatial resolution and tracking 
efficiency independent of (or even improving with) the track incident angle, and are not susceptible to aging. Already at the LHC design luminosity, the ATLAS MDT chambers are 
exposed to unprecedentedly high background rates of neutrons and $\gamma$-rays produced in interactions of the proton-proton collision products in the detector and shielding. 
At high-luminosity LHC (HL-LHC) 
as well as at future high-energy hadron colliders like FCC-hh, the background fluxes in the muon spectrometers are expected to increase roughly proportional to the luminosity by factors 
of up to 7 and 30, respectively. The maximum background flux MDT chambers are exposed to in ATLAS at the LHC design luminosity is 500~Hz/cm$^2$ while at FCC-hh the maximum 
expected rate in sMDT chambers is about 13~kHz/cm$^2$.

Small-diameter muon drift tube (sMDT) chambers~\cite{sMDT} with standard industrial aluminum tubes of 15~mm outer diameter and 0.4~mm wall thickness, 
i.e. with half of the MDT drift tube diameter, have been developed in order to
increase the spatial resolution and muon detection efficiency by more than an order of magnitude and to allow for larger numbers of drift tube layers within the available detector volume
in order to further improve the tracking resolution and efficiency, especially at high background rates. For ATLAS upgrades the gas composition and pressure and the gas gain of 20000 
as well as the front-end electronics chip layout are kept unchanged, making 15~mm tube diameter the optimum choice also with respect to mechanical precision and stability 
and electronics integration. Under these operating conditions, long-term irradiation tests have shown no aging up 9~C/cm charge accumulation on the wire, which is 15 times the requirement
for the ATLAS MDTs. For experiments beyond HL-LHC, further optimisation of the operating parameters is possible.

\section{Rate capability of small-diameter drift tube}

Due to their 4 times shorter maximum drift time (see Fig.~\ref{fig::spectrum}) and the twice higher granularity, sMDT drift tubes show 8 times smaller 
background occupancies compared to the MDT chambers. 
Deterioration of the spatial resolution due to radiation induced space charge fluctuations occurs only for drift distances larger than 7.5~mm, where the space-to-drift time relation
becomes non-linear, and is therefore completely eliminated in sMDT tubes.
The loss of gas gain due to shielding of the wire potential by the space charge and the resulting degradation of the resolution are suppressed proportional to the third power of the
inner tube radius and further due to the about $30\%$ lower average primary ionisation charge in sMDTs compared to MDTs (see Fig.~\ref{fig::gain}).
Because of the much shorter maximum drift time, the adjustable dead time of the MDT readout electronics (which for the MDTs is set to a nominal value of 820~ns, slightly above the maximum 
drift time of about 700~ns to prevent the detection of secondary ionization clusters) can be reduced to the minimum of 220~ns, just above the maximum drift time of the sMDT tubes of 175~ns. 
In this way the masking of muon hits
by preceding background pulses is vastly reduced increasing the muon detection efficiency defined as the probability to find a hit on the extrapolated 
muon track within 3 times the drift tube resolution ($3\sigma$ efficiency). 

The rate capability of the MDT and sMDT chambers has been extensively tested at the CERN Gamma Irradiation Facility using standard ATLAS MDT readout electronics with bipolar shaping.
In Figs.~\ref{fig:resolution} and \ref{fig:efficiency}, the measurements of the average spatial drift tube resolution and the 
$3\sigma$ muon efficiency as a function of the $\gamma$c background flux and hit rate, respectively, are compared to detailed simulations of drift tube and electronics response and agree
well with the expectations. The simulations also show that the vastly improved rate capability of the sMDTs can be further increased by employing readout electronics with fast baseline restoration (BLR) in order to prevent the 
signal pile-up effects, increased time jitter or even complete loss of muon signals overlapping with the bipolar undershoots of preceeding background
pulses, which deteriorate the resolution and efficiency. New fast readout electronics for the sMDT chambers with active baseline restoration is under development in order to fully exploit 
rate capability of the sMDTs. With such improved readout electronics it will likely become possible to reduce the electronics deadtime well below the maximum drift time of the
sMDT tubes. Muon detection efficiencies close to $100\%$ and spatial resolutions of better than $30\mu$m become possible with a minimum number of electronics channels.

\section{Conclusions}

Small-diameter drift-tube (sMDT) chambers provide efficient and robust, high precision muon tracking over large areas up to the highest background rates expected at future hadron colliders.
Experience with large-scale chamber production already exists from ongoing upgrades of the ATLAS muon spectrometer where sense wire positioning accuracies down to $5~\mu$m (rms) are 
routinely achieved~\cite{BMG}.

\end{document}